\documentclass[aps,pra,twocolumn,amsmath,amssymb,showpacs]{revtex4-1}
\usepackage{graphicx}
\usepackage{amsmath}
\usepackage{bm}

\newcommand{\be}{\begin{equation}}  
\newcommand{\ee}{\end{equation}}
\newcommand{\ba}{\begin{array}}
\newcommand{\ea}{\end{array}}
\newcommand{\bea}{\begin{eqnarray}}
\newcommand{\eea}{\end{eqnarray}}

\newcommand{\bra}{\langle}
\newcommand{\ket}{\rangle}

\newcommand{\nn}{\nonumber}

\newcommand{\aoop}{\overleftarrow}
\newcommand{\toop}{\overrightarrow}
\newcommand{\oms}{\omega_S}
\newcommand{\omr}{\omega_R}
\newcommand{\as}{A_S(\omega_S)}
\newcommand{\ar}{A_R(\omega_R)}

\newcommand{\adr}{A_R^{\dag}(\omega_R)}
\newcommand{\sinc}{{\rm sinc}}
\newcommand{\aso}{A_S(\omega)}

\begin{document}
\title{Apparent temperature: demystifying the relation between quantum coherence, correlations, and heat flows}

\author{C.L. Latune$^1$, I. Sinayskiy$^{1,2}$, F. Petruccione$^{1,2}$}
\affiliation{$^1$Quantum Research Group, School of Chemistry and Physics, University of
KwaZulu-Natal, Durban, KwaZulu-Natal, 4001, South Africa\\
$^2$National Institute for Theoretical Physics (NITheP), KwaZulu-Natal, 4001, South Africa}

\date{\today}
\begin{abstract}
Heat exchanges are the essence of Thermodynamics. In order to investigate non-equilibrium effects like quantum coherence and correlations in heat flows we introduce the concept of apparent temperature. Its definition is based on the expression of the heat flow between out-of-equilibrium quantum systems. Such apparent temperatures contain crucial information on the role and impact of correlations and coherence in heat exchanges. In particular, both behave as populations, affecting dramatically the population balance and therefore the apparent temperatures and the heat flows.
 We show how seminal results can be re-obtained, offering an interesting alternative point of view. We also present new predictions and suggest a simple experiment to test them.
 Our results show how quantum and non-equilibrium effects can be used advantageously, finding applications in quantum thermal machine designs and non-equilibrium thermodynamics but also in collective-effect phenomena.
 
\end{abstract}

\maketitle

\subsection*{Introduction}
It has been a long standing question whether new properties of thermodynamics can arise from quantum phenomena and assist thermodynamic tasks. Answering this question could unlock quantum advantages in thermodynamics and potentially revolutionise energy science \cite{Jaramillo_2016, Bengtsson_2017}. Such perspective alimented intense research around the exploitation of two genuine quantum characteristics, namely quantum coherence and correlations.    
The resource theory of coherence was developed in \cite{Baumgratz_2014, Lostaglio_2015, Winter_2016,Streltsov_2017, Morris_2018} and the role of coherence have been studied in thermal machines \cite{Scully_2011, Rahav_2012, Dorfman_2013, Brandner_2015, Uzdin_2015, Niedenzu_2015, Gelbwaser_2015, Leggio_2015b, Mitchison_2015, Killoran_2015, Uzdin_2016, Chen_2016, Su_2016,Turkpence_2016, Dag_2016, Mehta_2017, Dag_2018, Levy_2018, Xu_2018, Holubec_2018,Wertnik_2018} and work extraction \cite{Scully_2003, Aberg_2014, Li_2014, Llobet_2015, Korzekwa_2016, Kwon_2017, Vaccaro_2018}. Similarly, correlations were shown to provide the possibility to assist or enhance the performance of thermal machines \cite{Zhang_2007,Dillenschneider_2009,Wang_2009, Thomas_2011, Altintas_2014, Altintas_2015, Hardal_2015, Jaramillo_2016, Doyeux_2016, Dag_2016, Barrios_2017, Hardal_2017, Bengtsson_2017, Turkpence_2017, Dag_2018, Levy_2018, Gelbwaser_2018, Niedenzu_2018}, quantum battery charging \cite{Binder_2015, Campaioli_2017, Ferraro_2017, Le_2017}, work extraction \cite{Oppenheim_2002, Alicki_2013,Hovhannisyan_2013, Francica_2017, Manzano_2018}, energy transport \cite{Leggio_2015a}, and photovoltaic energy conversion \cite{Scully_2010, Svidzinsky_2011, Svidzinsky_2012,Creatore_2013}. The role of coherences and correlations in Landauer's Principle have also been studied \cite{Lorenzo_2015b, Kammerlander_2016}.   
  A broader understanding of such phenomena requires to go back to the essence of Thermodynamics, namely energy exchanges, the common ground of any thermal machines.   
Some papers already looked at the relation between energy flows and correlations in some particular models of heat engines \cite{Rahav_2012, Mitchison_2015} and in a pair of two-level systems with cascaded bath interaction \cite{Lorenzo_2015a}. In a different context, a series of interesting works \cite{Partovi_2008, Jennings_2010, Jevtic_2012, Micadei_2017, Henao_2018} investigate reversals of the natural heat flow (identified as the thermodynamic arrow of time) thanks to initial correlations {\it between} systems. 

Here, we develop a broad framework valid for any system, and introduce the concept of apparent temperature which characterises the thermodynamic behaviour of out-of-equilibrium quantum systems and determine the ongoing heat flows. Several effective temperatures have been defined previously in the literature, either based on the populations ratio of pairs of energy eigenstates \cite{Scovil_1959, Thomas_2011, Brunner_2012}, either based on the spectral density for non-thermal baths \cite{Alicki_2014}. The concept of virtual temperature together with virtual qubits \cite{Brunner_2012} brought great insights in the understanding of thermal machines \cite{Skrzypczyk_2015, Silva_2016, Erker_2017}. 
In the Discussion we come back in details on the main differences between apparent temperatures and virtual temperatures.
In particular, our concept of apparent temperature is based on the operators involved in the interaction between the systems. 
Interestingly, when one of the system can be described as a bath it can be shown that the apparent temperature coincides with the effective temperature mentioned in \cite{Alicki_2014, Alicki_2015}. 
 The introduction of the apparent temperature is motivated to use it to investigate the role of coherence and correlations in heat flows (which cannot be done with the previous concepts of temperature).   
 Its application to degenerate and many-body systems reveals how and why coherence and correlations present {\it within} the systems can impact and control the energy flows {\it between} them. Our results provide a deeper intuition and understanding of the interplay between correlations, coherence, and energy flows, essential for an efficient use of such resources.

\subsection*{Methods}
We consider two systems $S$ and $R$, possibly degenerate or composed of many (non-interacting) subsystems, with free Hamiltonian $H_S$ and $H_R$. They interact through the coupling Hamiltonian $H_{SR}=\lambda P_S P_R$, where $P_S$ and $P_R$ are observables of $S$ and $R$ respectively, and $\lambda$ characterises the strength of the coupling.   
For $S$ and $R$ initialised in {\it any states}, we want to investigate and characterise the energy exchange between them. 
This is a very challenging task in general. However, for short duration interactions, namely interaction time $\tau$ much smaller than the evolution timescale generated by the coupling, $|\lambda|^{-1}$ ($\hbar =1$), the energy exchange can be put in a suitable form. 
In order to observe a significative evolution, 
 we can consider repeated interactions between $R$ and $S$, assuming that $R$ is reinitialised in the same state or equivalently is replaced by an identical system in the same initial state. This corresponds to the well-known collisional model \cite{repeatedinteractions, Strasberg_2017}. Such a model is well adapted to describe certain types of experiments \cite{repeatedinteractions} and is very versatile 
 (describing successfully, beyond thermal baths, phenomena such as micromasers, feedbacks, Maxwell's demon, information reservoirs and repeated measurements \cite{Strasberg_2017}).
 To derive the reduced dynamics of $S$, we use the eigenoperator decomposition \cite{Petruccione_Book} of the observables $P_S$ and $P_R$, given by 
\be\label{eigenop}
P_X=\sum_{\omega_X\in {\cal E}_X} A_X(\omega_X),
\ee
 for $X=S,R$, where the eigenoperators (also called ladder operators) $A_X(\omega_X)$ satisfies $[H_X, A_X(\omega_X)] = -\omega_X A_X(\omega_X)$ and $A_X(-\omega_X) = A_X^{\dag}(\omega_X)$. 
 The sum runs over ${\cal E}_X$ denoting the ensemble of Bohr frequencies of the authorised transitions. The required condition $\tau \ll |\lambda|^{-1}$ mentioned above is equivalent to the assumption that the bath correlation time is much smaller than the timescale of the dissipation process when dealing with bath-induced dissipation \cite{Petruccione_Book, Cohen_Book}. In other words, the condition $\tau \ll |\lambda|^{-1}$ is equivalent to the Born and Markovian approximations for bath dissipation \cite{SM}. 
  We require furthermore that although $S$ and $R$ can be complex systems they still have discrete spectra and we assume that there exist only two kind of frequencies: the resonant or near resonant ones, satisfying $|\omega_S-\omega_R| \tau \ll 1$, while the others are far from resonance, meaning that $|\omega_S-\omega_R| \tau \gg 1$. This last condition is in fact not necessary, one can instead invoke the secular approximation \cite{SM}. However, the damping coefficients of the master equation takes a simpler form under the near-resonance assumption. 
  One should note that combined with the first condition, a detuning between $\omega_S$ and $\omega_R$ as large as $|\lambda|$ still satisfies the near-resonant condition. If two Bohr frequencies $\omega_S$ and $\omega_S'$ are quasi degenerate, namely $|\omega_S-\omega_S'| \tau \ll 1$, they are not resolved during the time interval $\tau$ and are therefore considered equal. The sum of their associated ladder operator is simply denoted by $A_S(\omega_S)$. With the same convention for $R$, we define ${\cal E}$ the intersection (up to $|\lambda|$) of ${\cal E}_S$ and ${\cal E}_R$.  
Under the above conditions, one can show (see details in Supplemental Material \cite{SM}) that the reduced dynamics of $S$ takes the form $ \rho_S(t+\tau) = (1+{\cal L}_{t,\tau} ) \rho_S(t)$,  where
  \bea
  {\cal L}_{t,\tau}\rho_S &=& - i \lambda \tau \sum_{\omega \in {\cal E}}[\aso,\rho_S] \bra A_R^{\dag}(\omega)\ket_{\rho_R^{Sc}(t)}e^{-i\omega t} \nn\\
 &&-\frac{\lambda^2\tau^2}{2}\sum_{\omega,\omega' \in {\cal E}} \bra A_R^{\dag}(\omega)A_R(\omega')\ket_{\rho_R^{Sc}(t)}e^{i(\omega'-\omega)t}\nn\\
 &&\!\!\!\!\!\!\!\!\!\! \times [\aso A_S^{\dag}(\omega') \rho_S - A_S^{\dag}(\omega')\rho_S\aso] + {\rm h.c.},
 \eea
where $\rho_S$ is the density matrix of $S$ in the interaction picture, and $\rho_R^{Sc}(t)$ is the density matrix of $R$ in the Schrodinger picture. 
Considering repeated interactions with $R$ (always prepared in the same state) at an average rate $r$, the coarse-grain derivative of $\rho_S$ is 
\bea\label{mer}
\dot{\rho}_S &=& \sum_{\omega \in {\cal E}}  \Gamma(\omega) \left(A_S(\omega)\rho_S A^{\dag}_S(\omega) - A_S^{\dag}(\omega) A_S(\omega)\rho_S\right)  \nn\\
&&+ {\rm h.c.} 
\eea
where $\Gamma(\omega) = \frac{r\lambda^2 \tau^2}{2}  {\rm Tr}[\rho_R A_R(\omega)A_R^{\dag}(\omega)]$. One should note that 
 the effect of coherences between levels of different energy cancel out on average due to the random phase of such coherences at each interaction \cite{SM}, 
  which is equivalent to stationary condition for baths (see \cite{Alicki_2014,Alicki_2015} and next paragraph). In particular, the unitary contribution, which is associated to work exchanges, is null on average. Interestingly, one recovers that work is performed only at the price of some ``order": if the phase of $R$ is not controlled and therefore random at each interaction, no work is performed. Conversely, the energy exchange associated to the second order term can be identified as heat. We will thus speak of heat exchanges in the remainder of the paper. \\

 The above dynamics \eqref{mer} described also the interaction
of the system $S$ with a bath $R$ in the Born-Markov
approximation  \cite{Petruccione_Book,Cohen_Book,SM}. The bath is not required to be in
a thermal state, however it must satisfy the condition $[\rho_R,H_R]=0$, sometimes called stationarity \cite{Alicki_2014, Alicki_2015}.
  Under such conditions the reduced dynamics of $S$ takes exactly the form \eqref{mer} but the function $\Gamma(\omega)$ has a different expression $\Gamma (\omega) = \lambda^2 \int_0^{\infty} ds e^{i\omega s}{\rm Tr}[\rho_R P^I_R(s) P_R]$,
where $P^I_R(s) = e^{iH_R s}P_R e^{-iH_R s}$ is the bath coupling operator in the interaction picture. According to the above considerations, in the remainder of the paper we will designate indifferently $R$ as a system interacting repeatedly with $S$ or as a bath. We introduce $G(\omega) := \Gamma(\omega) + \Gamma^*(\omega)$ the spectral density of $R$. Note that for a repeatedly interacting system the spectral density takes the simple expression $G(\omega)=  r\lambda^2 \tau^2  {\rm Tr}[\rho_R A_R(\omega)A_R^{\dag}(\omega)]$, whereas for a bath, $G (\omega) = \lambda^2 \int_{-\infty}^{\infty} ds e^{i\omega s}{\rm Tr}[\rho_R P^I_R(s) P_R]$.\\

We now derive the expression of the heat flow between $R$ and $S$ based on the dynamics \eqref{mer}. 
We define the internal energy of $S$ as $E_S := {\rm Tr} \rho_S H_S$ (invariant from the Schrodinger picture to the interaction picture).
 The heat flow from $R$ to $S$ is given by $\dot{Q}_{S/R} := \dot{E}_S = {\rm Tr}\dot{\rho}_S H_S$ \cite{Alicki_1979}. 
Using the expression \eqref{mer} of $\dot{\rho}_S$  the heat flow takes the simple form, 
\be
\dot{Q}_{S/R} = -\sum_{\omega \in {\cal E}}\omega G(\omega)  \langle A_S^{\dag}(\omega) A_S(\omega)\rangle_{\rho_S}.
\ee  
The bracket $\langle {\cal O} \rangle_{\rho_S}$ represents the expectation value of the operator ${\cal O}$ taken in the state $\rho_S$.

\subsection*{Results}
It is enlightening to consider first the situation where $\rho_R$ is a thermal state at temperature $T_R$. Then, the following identity holds (for both repeatedly interacting system and bath) \cite{SM}  (with $\hbar=1$, $k_{B} =1$), $G(-\omega) = e^{-\omega/T_R}G(\omega)$, and the heat flow can be rewritten in the interesting form
\bea\label{energyfl}
\dot{Q}_{S/R} &=& \sum_{\omega\geq 0}\omega G(\omega) \langle A_S(\omega) A_S^{\dag}(\omega) \rangle_{\rho_S} \nn\\
&&\hspace{1cm} \times \left[ e^{-\omega/T_R} - e^{-\omega/{\cal T}_S(\omega)}\right],
\eea
where we introduced the {\it time-dependent} parameter 
\be\label{apparenttemp}
{\cal T}_S(\omega) := \omega \left( \ln{\frac{\langle A_S(\omega) A_S^{\dag}(\omega) \rangle_{\rho_S}}{\langle A_S^{\dag}(\omega) A_S(\omega) \rangle_{\rho_S}}}\right)^{-1}.
\ee
When $S$ is in a thermal state at temperature $T_S$, the parameter ${\cal T}_S(\omega) = T_S$ for all $\omega$. 
However, when $S$ is in an arbitrary state, assuming $S$ and $R$ have only a single Bohr frequency in common, the parameter ${\cal T}_S(\omega)$ still plays the role of a temperature since it determines the sign of the heat flow: positive if ${\cal T}_S(\omega) \geq T_R$ and negative otherwise. For this reason we call ${\cal T}_S(\omega)$ apparent temperature: even if $S$ is not in a thermal state it appears at a temperature equal to ${\cal T}_S$ from the point of view of $R$. When $S$ and $R$ have multiple Bohr frequency in common, ${\cal T}_S(\omega)$ becomes the apparent temperature associated to the Bohr frequency $\omega$. It does not determine alone the sign of the total heat flow but determines the heat flow associated to the energy channel exchanging quanta of energy $\omega$. 
We can go further and define the same quantity for $R$. Then, we define, as for $S$, ${\cal T}_R(\omega):=\omega \left( \ln{\frac{\langle A_R(\omega) A_R^{\dag}(\omega) \rangle_{\rho_R}}{\langle A_R^{\dag}(\omega) A_R(\omega) \rangle_{\rho_R}}}\right)^{-1}$ the apparent (time-independent) temperature of $R$ associated to the Bohr frequency $\omega$. In the situation where $R$ is a bath we consider that $P_R$ can be decomposed in eigenoperators in a similar form as in \eqref{eigenop} (see \cite{SM}). Interestingly, one can show that the following identity 
\be\label{ident}
\frac{\langle A_R(\omega) A_R^{\dag}(\omega) \rangle_{\rho_R}}{\langle A_R^{\dag}(\omega) A_R(\omega) \rangle_{\rho_R}} = G(\omega)/G(-\omega),
\ee
holds even for baths \cite{SM} (and trivially if $R$ is a system interacting repeatedly with $S$). Using the above identity \eqref{ident} and  the parametrisation in terms of the apparent temperatures we can rewrite simply the heat flow between $S$ and $R$ as,
\bea\label{energyflow}
\dot{Q}_{S/R} &=& \sum_{\omega\geq 0}\omega G(\omega) \langle A_S(\omega) A_S^{\dag}(\omega) \rangle_{\rho_S} \nn\\
&&\hspace{1cm} \times \left[ e^{-\omega/{\cal T}_R(\omega)} - e^{-\omega/{\cal T}_S(\omega)}\right].
\eea
  Moreover, one should note that for $R$ the apparent temperatures can be expressed alternatively as ${\cal T}_R(\omega) = \omega[\log{G(\omega)/G(-\omega)}]^{-1}$ using the above identity \eqref{ident}. This corresponds to the effective temperature introduced for baths in \cite{Alicki_2014}. \\
 
  Assuming a single common Bohr frequency $\omega$, it is important to note that $S$ is led to thermalise to a state at a temperature {\it precisely equal to} ${\cal T}_R(\omega)$, re-enforcing the idea that ${\cal T}_R(\omega)$ corresponds to the apparent temperature of $R$ from the point of view of $S$. 
If $S$ and $R$ have multiple Bohr frequencies in common, then an apparent temperature  ${\cal T}_R(\omega)$ is associated to each Bohr frequency $\omega$ (taking note that ${\cal T}_R(-\omega) = {\cal T}_R(\omega)$).  
In general, the total energy flow between $R$ and $S$ is a sum of contributions from different channels of energy exchange $\omega$. Such channels correspond to the delocalised (i.e. from any level) absorption/emission of energy $\omega$ by $S$ and $R$. %
For each channel, the sign of the energy flow from $R$ to $S$ is given by the sign of ${\cal T}_R(\omega) - {\cal T}_S(\omega)$, as it would be for the energy flow between two thermal states at temperature $T={\cal T}_R(\omega)$ and $T={\cal T}_S(\omega)$, respectively.\\

\begin{figure}[t]
\centering
\includegraphics[width=0.45\textwidth]{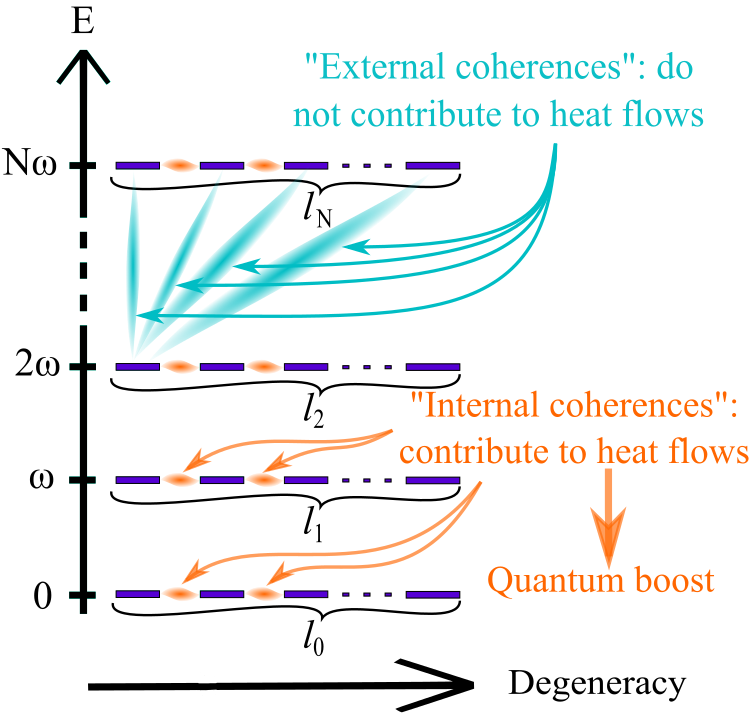}
\caption{Energy levels representation of a system of single Bohr frequency $\omega$. The energy is represented vertically and the degeneracy, horizontally.  ``Internal coherences" (coherences between degenerate states) sum up to populations to contribute to heat flows, whereas ``external coherences" (coherences between states of different energy) do not.}
\label{fig1}
\end{figure}

Importantly, one should keep in mind that the introduction of the concept of apparent temperature is based on the heat flow. Therefore, when there is no heat flow, there is no apparent temperature. In particular, no apparent temperature can be defined for non-interacting states (like dark states), neither it can for isolated systems. Furthermore, the same system in the same state can have different apparent temperatures associated to different interactions.  
Finally, the above formalism is not limited to the interaction with a bath or collisional model. As long as the dynamics can be described by Gorini-Kossakowski-Sudarshan-Lindblad (GKSL) master equation \cite{gkslme} the system can be assigned apparent temperatures of the form \eqref{apparenttemp}. The other system playing the role of bath or reservoir can be assigned apparent temperatures defined through the damping coefficients (the function $G(\omega)$) which might not in general admit a simple form as \eqref{apparenttemp}.   \\

{\bf The role of coherence}.
In this Section we investigate the relation between apparent temperatures and quantum coherence. 
We focus our analysis on situations where $S$ and $R$ have only a single Bohr frequency $\omega$ in common. Consequently, the energy flow between $S$ and $R$ is characterised by only one apparent temperature ${\cal T}_R$ and ${\cal T}_S$ for each system (in the remainder of the text we drop the explicit dependence on $\omega$ in ${\cal T}_S$, ${\cal T}_R$, and $A_R$).  
Since we assumed that $S$ and $R$ have only one frequency $\omega$ in common, only levels separated by $\omega$ are taking part in the dynamics (no space for spontaneous decay to other levels since $S$ and $R$ are isolated) so that $S$ and $R$ are effectively reduced to systems of equally separated levels. Thus, we consider that $S$ or $R$, designated by $X$ in the remainder of this Section, is a system of $(N+1)$ energy levels with degeneracy $l_n$ for each level $n$, $n \in [0,N]$ (see Fig.\ref{fig1}). In other words, the free Hamiltonian of $X$ can be written as $H_X = \sum_{n = 0}^N \sum_{g=1}^{l_n} n \omega |n,g\rangle \langle n,g|$, where the states $|n,g\rangle$ form an eigenbasis of $H_X$. 
The ladder operator $A_X$ is of the form (using the expression given earlier), 
\be\label{arsingle}
A_X = \sum_{n=1}^N \sum_{g = 1}^{l_{n-1}} \sum_{g'=1}^{l_n}\alpha_{n-1,n,g,g'}|n-1,g\rangle  \langle n,g'|,
\ee
with the coefficients $\alpha_{n-1,n,g,g'} := \langle n-1,g|P_X|n,g'\rangle$ describing the amplitude of probability transitions between levels. In simple situations the coefficients $\alpha$ are all equal and for simplicity we assume so in the remainder of the text as it does not change the nature of the results and simplify the expressions. The corresponding expression for the general situation can be found in the Supplemental Material \cite{SM}.
Inserting \eqref{arsingle} in \eqref{apparenttemp}, the apparent temperature is then equal to,
\be\label{cohapptemp}
{\cal T}_X = \omega \left( \ln{\frac{\sum_{n=1}^{N} l_n(\rho_{n-1} + c_{n-1})}{\sum_{n=1}^{N} l_{n-1}(\rho_n + c_n)}}\right)^{-1},
\ee
where $\rho_n := \sum_{g=1}^{l_n} \langle n,g|\rho_X|n,g\rangle$ is the sum of the populations of the degenerate levels of energy $n\omega$, 
 and $c_n  := \sum_{g \ne g' \in [1,l_n]} \langle n,g|\rho_X|n,g'\rangle$ 
 is the sum of the coherences between these degenerate energy levels. 
We can learn two things from \eqref{cohapptemp}. First, {\it coherences behave as populations} when it comes to heat exchange. This can change drastically the value of the apparent temperature and can be used to increase or decrease the apparent temperature. Denoting by ${\cal T}_X^0:= \omega \left( \ln{\frac{\sum_{n=1}^{N} l_n\rho_{n-1} }{\sum_{n=1}^{N} l_{n-1}\rho_n }}\right)^{-1}$ the apparent temperature without coherence, ${\cal T}_X$ is hotter than ${\cal T}_X^0$ (meaning $1/{\cal T}_X $ is smaller than $1/{\cal T}_X^0$, independently of their signs), if and only if $C^{+}\geq C^{-} e^{-\omega/{\cal T}_X^0}$, where $C^{+} := \sum_{n=1}^N l_{n-1}c_n$, and $C^{-} := \sum_{n=1}^N l_n c_{n-1}$. In the reverse situation ($C^{+}\leq C^{-} e^{-\omega/{\cal T}_X^0}$) ${\cal T}_X$ is colder than ${\cal T}_X^0$.

 Secondly, only coherences between degenerate levels affect the apparent temperature (which could be expected since the contributions to the dynamics from coherences between levels of different energy cancels out).  
In a different context the authors of \cite{Kwon_2017} also observed a similar dichotomy: coherences between degenerate states, called ``internal coherences", affect the work extraction whereas coherences between states of different energy, called ``external coherence", do not. It would be interesting to investigate further how these two phenomena are related.
As a consequence, coherences in non-degenerate systems cannot affect energy flows (unless by taking part to a many-body system, see next Section), but interesting effects of non-thermal population distributions can still persist (see \cite{refri}). \\

\begin{figure}[t]
\centering
\includegraphics[width=0.46\textwidth]{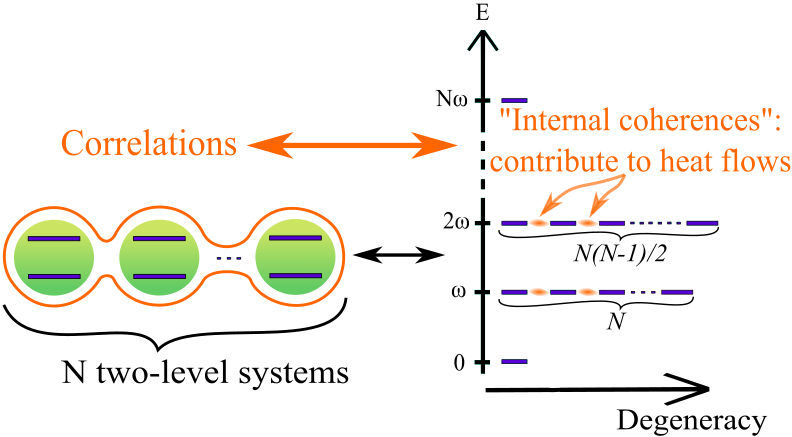}
\caption{Example of a cloud of $N$ two-level systems. The cloud (left-hand side) can be described as a single system with degenerate energy levels (right-hand side). The correlations correspond to internal coherences in the energy levels representation, revealing an equivalence between them.}
\label{fig2}
\end{figure}

{\bf The role of correlations}. In this paragraph we investigate the relation between the apparent temperature and correlations.
We consider $X$ (still designating either $S$ or $R$) as an ensemble of $N$ non-interacting subsystems (not necessarily identical) of {\it same} Bohr frequency $\omega$. In order to follow the dynamics described by \eqref{mer} one important condition is that all the $N$ subsystems have to experience the same environment, which usually requires confinement in a volume smaller than the typical variation length scale of that environment, like the electromagnetic emission wavelength in the Dicke model \cite{Dicke_1954,Gross_1982}.
  Upon the above conditions the $N$ subsystems interact {\it collectively} through the collective ladder operator $A_X = \sum_{n=1}^{N} A_n$, where $A_n$ is the ladder operator of the subsystem $n$ (with the same properties as above). 
Using the expression \eqref{apparenttemp}, the apparent temperature is,
\bea\label{correlationtemp}
{\cal T}_X &=&  \omega \left( \ln{\frac{ \sum_{i=1}^m \langle A_i A_i^{\dag}\rangle_{\rho_X}+c }{\sum_{i=1}^m \langle A_i^{\dag} A_i \rangle_{\rho_X} + c}}\right)^{-1},
\eea
where $c:= c_{\rm cor} +c_{\rm coh}$ is the sum of the correlations between the $N$ subsystems, $c_{\rm cor}=\sum_{i\ne j=1}^m [\langle A_i A_j^{\dag}\rangle_{\rho_X}-\langle A_i\rangle_{\rho_X} \bra A_j^{\dag}\rangle_{\rho_X}]$, and products of {\it local} cross coherences $c_{\rm coh} =\sum_{i\ne j=1}^m \langle A_i\rangle_{\rho_X} \bra A_j^{\dag}\rangle_{\rho_X}$.
  As previously, {\it correlations and local cross coherences act as populations}. Their impact on the apparent temperature, and consequently on the energy flow, can also be significant (see also next Sections). In particular, one can show that the correlations and local cross coherences let ${\cal T}_X$ hotter (colder) if and only if $c$ is positive (negative), assuming the apparent temperature with $c=0$ is positive (or the individual subsystems are not in inverted-population states), otherwise it is the contrary \cite{SM}. Interestingly, the resulting effects on the apparent temperature do not distinguish genuine correlations from product of cross coherences: what is important is having $c\ne 0$ which can come either from local coherences or correlations.
Taking forward the comparison with coherence in a single system, one can recast the states of the $N$ subsystems in the same form as a single system with degenerate energy levels, see Fig.\ref{fig2} (exemplified with $N$ two-level systems). Then, one can see that in this new picture $c$ is precisely the sum of the {\it global} coherences between degenerate levels of the global systems $S$. In this sense this second situation of many-body systems is a particular case of the first one studied in the previous Section. This provides a simple way to identify in a many-body system which correlations and product of local coherences participate to the heat flow and to the apparent temperature. 
 Although coherence is a genuine quantum feature, correlations not necessarily. In particular, there is no distinction between classical or quantum correlations in the effect produced on the apparent temperature and the heat flow. We recover the known observation \cite{Dillenschneider_2009, Jennings_2010, Dag_2018} that classical correlations can lead to the same effects as quantum correlations (despite the fact that quantum correlations can be stronger and therefore lead to a larger effects).  \\

{\bf Illustrative examples from the literature}.
In \cite{Dillenschneider_2009}, a photonic engine is powered by a pair of two-level atoms. The pair plays the role of the system $R$ repeatedly interacting with an optical cavity field (the working media of the engine) via the Tavis-Cummings coupling, $H_{\rm int} = - g (a S^{\dag} + a^{\dag} S^{-})$, with
coupling constant $g$. The collective ladder operator $S^{-}$ is given by $S^{-}  = |g\rangle_1\langle e| + |g\rangle_2\langle e|$, where $|e\rangle_i$ and $|g\rangle_i$ are the excited and ground states of the atom $i=1,2$, respectively, and  $S^{\dag}$ is given by the hermitian conjugate of $S^{-}$. 
The optical cavity mode is brought to equilibrium through repeated interactions with the pair of atoms. One central result of \cite{Dillenschneider_2009} is that the equilibrium temperature of the optical cavity is shown to be smaller when the pair of atoms is prepared in thermally entangled states. This is exploited to increase the efficiency of the engine. \\
Applying the expression \eqref{correlationtemp} with $A_i =  |g\rangle_i\langle e|$, $i=1,2$, the apparent temperature of the pair of atoms is 
\be\label{temppair}
{\cal T}_{\rm at} = \omega \left( \ln{\frac{\rho_g +\rho_d + \rho_{nd}}{\rho_e +\rho_d +\rho_{nd}}}\right)^{-1},
\ee
where $\rho_g := \langle g,g|\rho|g,g\rangle$, $\rho_e := \langle e,e|\rho|e,e\rangle$, $\rho_{d} := (\langle e,g|\rho|g,e\rangle + \langle g,e|\rho|e,g\rangle)/2$, and $\rho_{nd} := (\langle g,e|\rho|e,g\rangle + \langle e,g|\rho|g,e\rangle)/2$. The apparent temperature \eqref{temppair} is precisely the equilibrium temperature of the cavity mode found in \cite{Dillenschneider_2009} (where $\langle e,g|\rho|g,e\rangle = \langle g,e|\rho|e,g\rangle$ and $\langle g,e|\rho|e,g\rangle = \langle e,g|\rho|g,e\rangle$ due to the symmetric states used therein), and  in \cite{Dag_2018} (when neglecting the cavity damping). This comes as a confirmation of our claim: correlations alter the heat flows bringing the cavity mode to {\it thermalise} at the altered apparent temperature \eqref{temppair}. Furthermore, the correlation term for the kind of states considered in \cite{Dillenschneider_2009} is $c=-\frac{1}{Z}\sinh{\lambda \beta}$, where $\lambda$ is the coupling strength between the atoms, $\beta$ is the underlying inverse temperature of the atoms, and $Z$ is the partition function. Then $c \leq 0$, confirming our prediction that negative correlations decrease the apparent temperature. \\

Our second illustration concerns a three-level system in ``$\Lambda$-configuration" as used in \cite{Scully_2002, Scully_2003}. The lower states $|b\rangle$ and $|c\rangle$ are (almost) degenerate and separated from the exited state $|a\rangle$ by the transition energy $\omega$. The three-level system interacts repeatedly with  an electromagnetic field of a cavity mode through the Jaynes-Cummings Hamiltonian,
$H_{RB} = \lambda (|b\rangle \langle a| + |c\rangle \langle a|) d^{\dag} + {\rm h.c.}$,
where $d^{\dag}$ is the creation operator of the cavity mode.
  The apparent temperature of the three-level system obtained from \eqref{cohapptemp} with $A_R= |b\rangle \langle a| + |c\rangle \langle a| $ is
\bea\label{tempphaseo}
{\cal T}_{\rm ph} =\omega \left( \ln{\frac{\rho_{bb}+\rho_{cc}+\rho_{bc}+\rho_{cb}}{  2\rho_{aa}}}\right)^{-1},
\eea
where $\rho_{xy}:=\langle x|\rho_R|y\rangle$, $x,y= a,b,c$, and is increased by {\it negative} coherences (i.e. $c_0=\rho_{bc}+\rho_{cb}\leq0$). In \cite{Scully_2002, Scully_2003, Dag_2018} the authors consider an ensemble of three-level atoms prepared in coherent states, the so-called phaseonium, interacting with an optical cavity mode. Their central result is that the steady state temperature of the cavity mode when reaching equilibrium with the phaseonium is affected by the existence of coherence between the state $|b\rangle$ and $|c\rangle$ of the atoms. By preparing such coherence with {\it negative} real part the steady state temperature of the cavity is significantly increased yielding an increase of the engine efficiency.
This steady state temperature is precisely \eqref{tempphaseo} (see \cite{SM}). 
Again, this confirms the predictions of our framework: even though the atoms are not in a thermal state, they appear to the cavity field at the apparent temperature \eqref{tempphaseo} (increased by negative coherences), and the cavity field eventually thermalises at this temperature.  

Additionally, the three-level system in ``$\Lambda$-configuration" is also used in lasing without inversion \cite{Scully_1989}. One can find straightforwardly the essence of such phenomenon: the inversion of population 
usually required for lasers can be replaced by negative apparent temperature of the atoms since the effects of apparent temperatures are the same as usual temperatures. 
Then, as above, coherences prepared such that $\rho_{bc}+\rho_{cb} < 0$ 
can assist atoms reaching an apparent negative temperature, or apparent inversion, realising lasing without inversion).\\

{\bf Applications and Predictions}.
Now we derive one of the curious consequences of the above framework.
We consider a cloud of $N$ two-level atoms interacting with the surrounding electromagnetic field assumed to form a thermal bath at temperature $T_B = \beta_B^{-1}$. 
According to \cite{Gross_1982}, if the $N$ atoms are confined in a volume of dimensions smaller than $\omega/c$ the electromagnetic wavelength corresponding to the atomic transition, the reduced dynamics of the cloud of two-level atoms with parallel polarisation of the electric dipole is (in the interaction picture)
\bea\label{ind}
\dot{\rho}_S^I &=& -i\Omega_L \sum_{i=1}^N [\sigma_i^{+}\sigma_i^{-},\rho_S^I]-i\sum_{i > j}\Omega_{i,j}[\sigma_i^{+}\sigma_{j}^{-}+\sigma_i^{-}\sigma_{j}^{+},\rho_S^I] \nn\\
&&+ g [n(\omega) +1] (2S^-\rho_S^I S^+ -S^+S^-\rho_S^I - \rho_S^IS^+S^-)\nn\\
&&+ g n(\omega) (2S^+\rho_S^I S^- -S^-S^+\rho_S^I - \rho_S^IS^-S^+),\nn\\
\eea
where $n(\omega)$ is the bath mean excitation number at the frequency $\omega$, $\sigma_i^+$ and $\sigma_i^-$ are the ladder operators of the atom $i$ at the position $\vec r_i$, $S^+ = \sum_{i=1}^N\sigma_i^+$ and $S^-= \sum_{i=1}^N\sigma_i^- $ are the {\it collective} ladder operators of the cloud of atoms, $\Omega_L$ is the Lamb shift, $g= \frac{d^2\omega^3}{6\pi c^3\hbar \epsilon_0}$ with $d$ is the electric dipole of one atom (both atoms are identical), $c$ is the vacuum light velocity and $\epsilon_0$ is the vacuum permeability. The terms proportional to $\Omega_{i,j}=\frac{d^2}{4\pi\epsilon_0r_{ij}^3}\left[1-\frac{3(\vec{\epsilon}_a.\vec r_{ij})^2}{r_{ij}^2}\right]$ correspond to the Van der Waals interaction between the two atoms where $\vec r_{ij}=\vec r_i -\vec r_j$, and $\vec \epsilon_a$ is the polarisation of the electric dipole (assumed to be the same for all atoms). The above master equation expresses the indistinguishability of each atom from the point of view of the bath. One should note that since the interaction term conserves the total energy of the cloud, the expression of apparent temperature of the cloud is still given by \eqref{correlationtemp}. By contrast, when the $N$ atoms are distinguishable, either because they are separated by a distance larger than $\omega/c$ or because their polarisation is orthogonal, the reduced dynamics of the cloud is 
\bea\label{dis}
\dot{\rho}_S^I &=& -i\Omega_L \sum_{i=1}^N [\sigma_i^{+}\sigma_i^{-},\rho_S^I] \nn\\
&&+ g [n(\omega) +1]\sum_{i=1}^N (2\sigma_i^-\rho_S^I \sigma_i^+ -\sigma_i^+\sigma_i^-\rho_S^I - \rho_S^I\sigma_i^+\sigma_i^-)\nn\\
&&+ g n(\omega) \sum_{i=1}^N(2\sigma_i^+\rho_S^I \sigma_i^- -\sigma_i^-\sigma_i^+\rho_S^I - \rho_S^I\sigma_i^-\sigma_i^+),\nn\\
\eea
which corresponds to $N$ atoms interacting independently with the same bath. In particular the  apparent temperature of the cloud is 
\be\label{distemp}
{\cal T}_{\rm dis} =\omega\left(\log \frac{\sum_{i=1,2}\bra \sigma_i^{-}\sigma_i^{+}\ket}{\sum_{i=1,2}\bra\sigma_i^{+}\sigma_i^{-}\ket}\right)^{-1},
\ee
 which corresponds to the expression \eqref{correlationtemp} without the non-diagonal terms. 

\begin{figure}[t]
\centering
\includegraphics[width=0.46\textwidth]{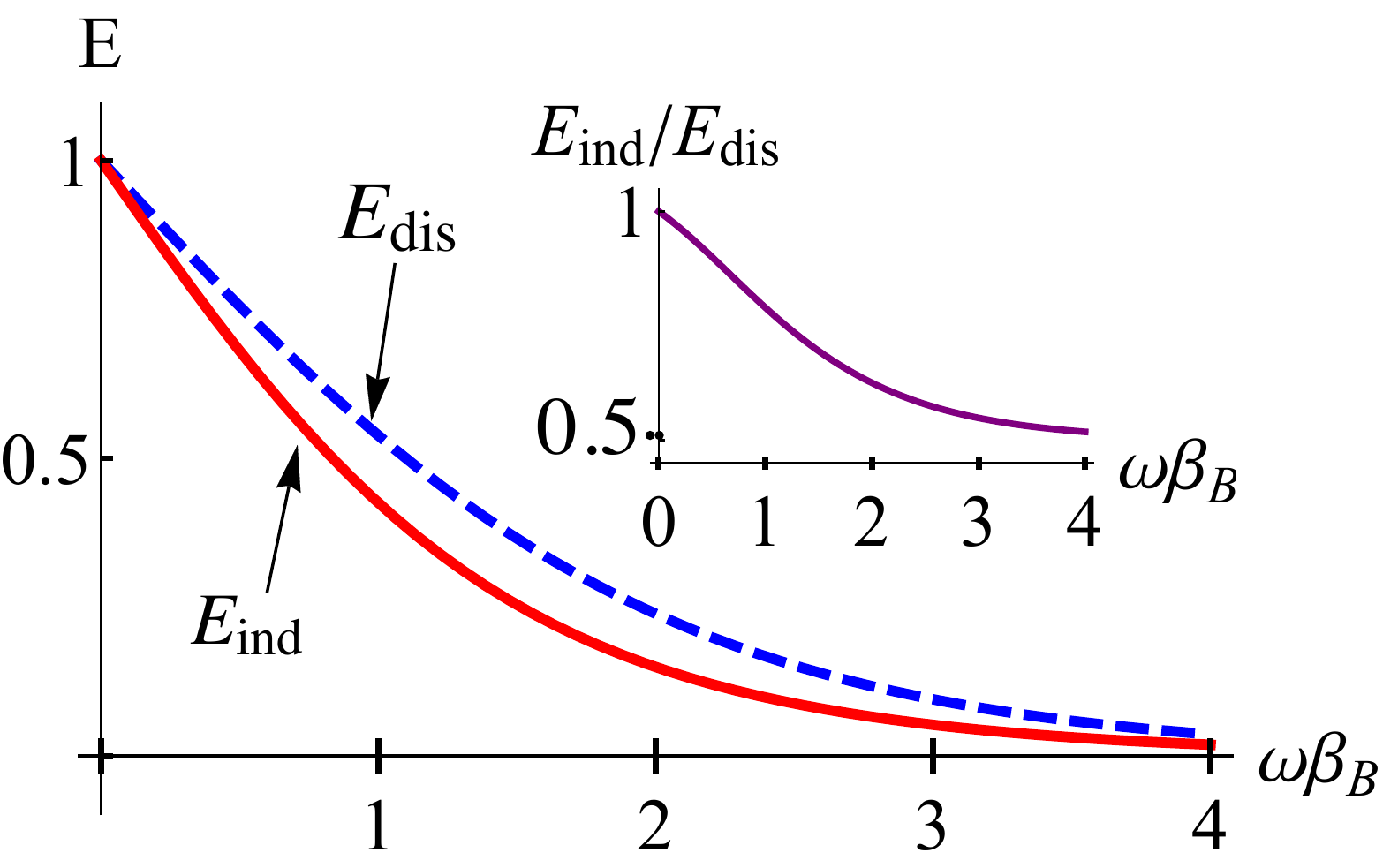}
\caption{Graph of the steady energy $E_{\rm dis}/\omega$ (blue, dashed) and $E_{\rm ind}/\omega$ (red) in function of $\omega\beta_B$. The insert is the curve of $E_{\rm ind}/E_{\rm dis}$ as a function of $\omega\beta_B$.}
\label{graph}
\end{figure}

The above two master equations describe the process of absorption of a bath excitation when the atoms are distinguishable \eqref{dis} and indistinguishable \eqref{ind} to the bath. In particular, assuming that both atoms are in the ground state at the instant of time $t_0$, one can see from the master equation \eqref{dis} that at $t_0 + \delta t$ (with $\delta t \ll |\dot \rho_S^I|\simeq g n(\omega)$) the distinguishable cloud has a probability $2Ngn(\omega)\delta t$ to absorb a bath excitation and to be projected into the mixed state $(|10...0\ket\bra0...01| + ...+ |00...1\ket\bra1...00|)/N$. During the excitation absorption the apparent temperature jumps from zero to ${\cal T}_{\rm dis} = \frac{\omega}{\ln{(N-1)}}$ (obtained from \eqref{distemp}). Then, for a large number of atoms the apparent temperature is almost not modified by the absorption of one bath excitation, as expected. 
 By contrast, the indistinguishable cloud has the same probability to absorb a bath excitation and to be projected onto the the coherent superposition $(|10...0\ket+...+|00...1\ket)/\sqrt{N}$ while the apparent temperature jumps from 0 to ${\cal T}_{\rm ind} = \omega/\log [2(1-1/N)] \simeq 1.5 \omega$ (for large $N$). This is strikingly different from the previous situation. First the apparent temperature does not go to zero for large samples, but also the impact on the apparent temperature of an excitation absorption tends to be independent of the number of atoms for large $N$. It means that localised and delocalised excitations affect systems in profoundly different ways.
Such surprising predictions could be verified by looking at the steady state. Indeed, although it should be difficult to observe directly this phenomenon, it should have a strong impact on the thermalisation process: each time the ensemble of $N$ atoms absorbs (collectively) an excitation from the bath, the apparent temperature of the the cloud rises considerably. It is then expected that the apparent temperature reaches quickly its steady state value ${\cal T}_{\rm ind} = T_B$ after having absorbed very few bath excitations: an ``apparent thermalisation''. Therefore, the steady state energy should be  
   much lower than for a thermal state (corresponding to a cloud interacting distinguishably with the bath). In order to realise the experiment the breakdown of symmetry induced by interactions between atoms has to be avoided by arranging the atoms in ring configuration \cite{Gross_1982} or by using a cavity to select the electromagnetic modes involved in the dissipation process. However, for a cloud of only two atoms such experimental difficulties are avoided naturally \cite{Gross_1982} and the desired effect is still present. 
   The steady state corresponding to the dynamics \eqref{ind} for a pair of atoms initially in the ground state is \cite{SM} 
   \bea
   \rho_{\rm ind}^{\infty} = Z_{\rm ind}^{-1} \Big(e^{-2\omega\beta_B}|11\ket\bra 11|&+&e^{-\omega\beta_B}|\psi_+\ket\bra\psi_+| \nn\\
   &+& |00\ket\bra00|\Big),
\eea
with $Z_{\rm ind}:=(1+e^{-\omega\beta_B} +e^{-2\omega\beta_B})$ and $|\psi_+\ket := (|01\ket+|10\ket)/\sqrt2$. The energy associated to $\rho_{\rm ind}^{\infty}$ is $E_{\rm ind}: = \omega e^{-\omega\beta_B}(1+2e^{-\omega\beta_B})/Z_{\rm ind}$ whereas the steady state energy of the equilibrium thermal state $\rho_{dis}^{\infty}  :=Z_{\beta_B}^{-1} e^{-H_S\beta_B}$, with $Z_{\beta_B} = {\rm Tr} e^{-H_S \beta_B}$, reached by the distinguishable pair of atoms is $E_{\rm dis}:= 2\omega e^{-\omega\beta_B}/(1+e^{-\omega\beta_B})$. In Fig. \ref{graph} we plot the graph of the steady energy $E_{\rm ind}$ and $E_{\rm dis}$ as a function of the bath temperature $\omega\beta_B$. The inset represents the curve $E_{\rm ind}/E_{\rm dis}$ as a function of $\omega\beta_B$, which goes down to $0.5$: indistinguishable pair of atoms has an apparent temperature up to $50\%$ smaller than a distinguishable pair. Such effect should be within reach of current experiments in cold atoms and superradiance. 

\subsection*{Discussion}
As announced in the introduction we come back now on the concept of virtual temperature introduced in \cite{Brunner_2012}. The virtual temperature associates a temperature to any pair of energy levels $|\epsilon_i\ket$ and $|\epsilon_j\ket$ based on their energy difference $\epsilon_i - \epsilon_j$ and on their respective populations $p_i$ and $p_j$,
\be
T_v= \frac{1}{\epsilon_i-\epsilon_j} [\log{p_j/p_i}]^{-1}.
\ee
Thanks to the associated notion of virtual qubit \cite{Brunner_2012} the application of virtual temperature goes beyond two-level systems. 
Then, one can transform the virtual temperature of a virtual qubit to a ``real" temperature by engineering adequate interactions (for instance two thermal baths put in thermal contact with virtual qubit of interest) \cite{Brunner_2012,Skrzypczyk_2015, Silva_2016}. It appears to be a precious tool to analyse and engineer thermal machines \cite{Skrzypczyk_2015, Silva_2016, Erker_2017}. However, the virtual temperature presents a series of difference with apparent temperature, beginning with the context and objective of their use and definition. The virtual temperature has no direct link with energy flow. If one is interested in energy flow, one has to work out this link and even for simple situations like the interaction between a harmonic oscillator and a bath it requires an infinite number of virtual temperatures. 
On the other hand, the apparent temperature definition is based on the heat flow so that the apparent temperature is intrinsically related to the heat flow. In particular, no apparent temperature can be defined where there is no heat flow. Interestingly, the apparent temperature takes a simple form expressed in terms of the coupling operators. Only in the special situation of a single two-level system the apparent temperature is determined by the ratio of the excited level population by the ground level population, and therefore equivalent to the virtual temperature. In any other system the apparent temperature is not equal and not equivalent to the virtual temperature. For instance a harmonic oscillator undergoing a dissipation process described by \eqref{mer} is simply described (from the point of view of heat flows) by a unique apparent temperature, whereas it requires an infinite number of virtual temperatures as already mentioned.
Finally and more importantly, virtual temperatures do not accommodate coherence. As a consequence, the virtual temperature fails to describe the important role of coherence and correlation in heat flows, which is precisely the aim of this paper.
Recapping, rather than a generalisation of virtual temperature, we see the apparent temperature as an alternative concept aiming principally to describe heat flows between arbitrary quantum systems in arbitrary states, whereas the virtual temperature is more suitable and conceived to study and engineer thermal machines in non-degenerate systems of limited dimensions. 

\subsection*{Conclusions}

Throughout this study we show that the concept of apparent temperature captures well the thermodynamic behaviour of out-of-equilibrium quantum systems even when usual quantities like temperatures or out-of-equilibrium free energy are not defined. In particular, it determines the instantaneous Markovian heat flows between quantum systems.
 Its application to complex systems such as degenerate or many-body systems demystifies the tight relations between correlations and coherences within the systems and heat flows. 
 Our results reveal that coherences between degenerate levels of a system behave as populations when it comes to heat exchanges with external systems. In particular, when the system of interest is a many-body system, both correlations between subsystems and product of local coherences which can be identified as correlations between degenerate levels of the global system behave as populations. 
 This is a surprising effect since usually only populations participate to heat flows. 
  This apparent contradiction comes from the fact that heat exchanges are determined by the expectation values of the operators $ A_X A_X^{\dag}$ and $ A_X^{\dag} A_X $, different from the free Hamiltonian $H_X$. Then, whereas coherences between degenerate levels are off-diagonal elements in the natural eigenbasis of $H_X$ they can contribute to the expectation values of $A_X A_X^{\dag}$ and $A_X^{\dag} A_X$ (becoming diagonal elements in the eigenbasis of  $A_X A_X^{\dag}$ and $A_X^{\dag} A_X$). The resulting effects on the heat flow are captured by the apparent temperature. 
 As equivalent explanation, GKSL master equations for non-degenerate systems produce independent dynamics for population and coherences. However, for degenerate systems the dynamics of populations and coherence between degenerate levels is mixed together. As a result, coherences between degenerate levels affect the population's dynamics and therefore the heat flows and the apparent temperature.
This is in sharp contrast with classical thermodynamics. Then, given that heat flows are at the heart of Thermodynamics, quantum Thermodynamics could be expected to be strikingly different from its classical counterpart. Moreover, such effects are universal (not limited to a particular system, configuration, or platform) and should be testable in relatively simple experiments monitoring the energy of indistinguishable pair of two-level atoms. It can have important consequences in equilibration processes but also in quantum thermal machines \cite{refri}, out-of-equilibrium thermodynamic problems, phenomena involving collective effects like superradiance or even light-harvesting complexes \cite{Chin_2013} and related photocells \cite{Scully_2011,Creatore_2013}.

\begin{acknowledgments}
This  work  is  based  upon  research  supported  by  the
South  African  Research  Chair  Initiative  of  the  Department  of  Science  and  Technology  and  National  Research Foundation.
\end{acknowledgments}

\appendix
\begin{widetext}
\section*{Supplemental Material}

 \section{Repeated interactions/Collisional model}
 Derivations of Markovian master equations based on the collisional model can be found in \cite{repeatedinteractions, Strasberg_2017}. In the following we re-derive in simple term such Markovian master equation. At the instant of time $t$ the system $S$, found in the state $\rho_S^{Sc}(t)$, interacts for a duration $\tau$ with the system $R$ in the state $\rho_R^{Sc}(t)$. The superscript $Sc$ stands for Schrodinger picture. The evolution of $SR$ between $t$ and $t+\tau$ is given by
 \bea
 \rho_{SR}^{\rm Sc}(t+\tau) &=&  e^{-i\tau H_{\rm Tot}} \rho_S^{\rm Sc}(t)\rho^{Sc}_R(t) e^{i\tau H_{\rm Tot}}\nn\\
 &=& e^{-i\tau (H_S+H_R)} e^{-i \toop T \int_0^{\tau} {\rm d}u \tilde{H}_{SR}(u)} \rho_S^{\rm Sc}(t) \rho^{Sc}_R(t)  e^{i\aoop T\int_0^{\tau} {\rm d}u \tilde{H}_{SR}(u)}  e^{i\tau (H_{S}+H_R)}
  \eea
  where $H_{\rm Tot} = H_S +H_R +H_{SR}$ is the total Hamiltonian, $\toop T$ and $\aoop T$ are respectively the time ordering and anti-chronological ordering operators, and $\tilde{H}_{SR}(u) = e^{i u(H_S+H_R)} H_{SR}e^{-iu(H_S+H_R)}$ stands for the interaction picture of $H_{SR}$. 
  Assuming the duration of the interaction $\tau$ is much smaller than the evolution timescale induced by the coupling $H_{SR}$, namely $\tau \ll |\lambda|^{-1}$, we can expand the time ordered integrals which yields
  \bea\label{expansion} 
  e^{i\tau (H_S +H_R)} \rho_{SR}^{Sc}(t+\tau) e^{-i\tau(H_S+H_R)} &=& \rho_S^{Sc}(t)\rho_R^{Sc}(t) - i \int_0^{\tau} {\rm d}u [\tilde{H}_{SR}(u), \rho_S^{Sc}(t)\rho_R^{Sc}(t)] \nn\\
 && - \int_0^{\tau}{\rm d}u_1 \int_0^{u_1}{\rm d}u_2 [\tilde{H}_{SR}(u_1),[\tilde{H}_{SR}(u_2),\rho_S^{Sc}(t)\rho_R^{Sc}(t)]] + {\cal O}(\tau^3|\lambda|^3),
  \eea
  and for the reduced dynamics of $S$ (tracing out $R$),
  \bea\label{reduced}
   e^{i\tau H_S} \rho_{S}^{Sc}(t+\tau) e^{-i\tau H_S} &=& \rho_S^{Sc}(t) - i \int_0^{\tau} {\rm d}u {\rm Tr}_R [\tilde{H}_{SR}(u), \rho_S^{Sc}(t)\rho_R^{Sc}(t)] \nn\\
 && - \int_0^{\tau}{\rm d}u_1 \int_0^{u_1}{\rm d}u_2 {\rm Tr}_R[\tilde{H}_{SR}(u_1),[\tilde{H}_{SR}(u_2),\rho_S^{Sc}(t)\rho_R^{Sc}(t)]] + {\cal O}(\tau^3|\lambda|^3).
  \eea
One can easily show that if we consider the non-truncated expansion, the $n^{\rm th}$ term is at most at most of order $(|\lambda|\tau)^n/n!$, so that we can safely neglect the higher orders as long as $\tau \ll |\lambda|^{-1}$.   
  To provide a simpler expression for the reduce dynamics of $S$ we rewrite the coupling Hamiltonian $H_{SR} = \lambda P_SP_R$ in terms of the eigenoperators of $S$ and $R$ (also called ladder operators) $\as$ and $\ar$ which verify the relations $[H_X, A_X(\omega_X)] = -\omega_X A_X(\omega_X)$ ($\hbar=1$) and $A_X(-\omega_X) = A_X^{\dag}(\omega_X)$ for $X=S,R$. Such eigenoperators can be obtained from the operators $P_S$ and $P_R$ through \cite{Petruccione_Book} $A_X(\omega_X):= \sum_{\epsilon' -\epsilon = \omega_X} \pi_{\epsilon} P_X \pi_{\epsilon'}$, $X=S,R$,
where $\pi_{\epsilon}$ is the projector onto the eigenspace associated with the eigenvalue $\epsilon$ of $H_X$. Then, the operator $P_X$ can be written as $P_X =\sum_{\omega_X\in {\cal E}_X} A_X(\omega_X) =\sum_{\omega_X \in {\cal E}_X} A_X^{\dag}(\omega_X)= \sum_{\omega_X > 0} [A_X(\omega_X) + A_X^{\dag}(\omega_X)]$, where ${\cal E}_X$ denotes the ensemble of the Bohr frequencies of the system $X=S,R$. The coupling Hamiltonian can be re-written as
  \be
  H_{SR} = \lambda P_S P_R = \lambda \sum_{\oms \in {\cal E}_S, \omr\in {\cal E}_R} \as \ar = \lambda \sum_{\oms \in {\cal E}_S, \omr \in {\cal E}_R} \as \adr,
  \ee
  and in the interaction picture it is simply $\tilde{H}_{SR}(u) =\lambda \sum_{\oms \in {\cal E}_S,\omr\in {\cal E}_R} e^{i(\omr -\oms)u} \as \adr$.
  The term of first order in the expansion \eqref{reduced} becomes
  \bea
   - i \int_0^{\tau} {\rm d}u {\rm Tr}_R[\tilde{H}_{SR}(u), \rho_S^{Sc}(t)\rho_R^{Sc}(t)] &=& - i \lambda \sum_{\oms\in {\cal E}_S}[\as,\rho_S^{Sc}(t)]\sum_{\omr\in {\cal E}_R} {\rm Tr}_R \adr \rho_R^{Sc}(t) \int_0^{\tau} {\rm d} u e^{i(\omr-\oms)u} .
   \eea
   The integral is equal to $\int_0^{\tau} {\rm d} u e^{i(\omr-\oms)u} = \tau e^{i(\omr-\oms)\tau/2}  \sinc(\omr-\oms)\tau/2$ which is approximately equal to $\tau$ when $|\omr-\oms| \tau \ll 1$ and much smaller than $\tau$ when $|\omr-\oms|\tau \gg1$. Although we allow $S$ and $R$ to be complex systems we can assume that they have discrete spectra, and that there exists only two kinds of frequencies, those in resonance or near-resonance satisfying $|\omr-\oms| \tau \ll 1$ and those far from resonance such that $|\omr-\oms| \tau \gg 1$. Furthermore, all quasi degenerate frequencies $\omega_R$ and $\omega_R'$ such that $|\omr-\omega_R'| \tau \ll 1$ are considered equal since they are not resolved during the time interval $\tau$. Consequently, $\ar$ turns to designate the sum of the ladder operators of quasi degenerate frequencies $\omr$, 
\be
\ar \rightarrow \sum_{\omega_R';|\omr-\omega_R'| \tau \ll 1}A_R(\omega_R').
\ee
 We adopt the same convention for the quasi degenerate Bohr frequencies of $S$ and for $A_S(\omega_S)$. Under such conventions, for any $\oms \in {\cal E}_S$, there exists at most one unique frequency $\omr \in {\cal E}_R$ such that $\omr \simeq \oms$ (understand $|\omr - \oms|\tau \ll1$). We denote the intersection of the two ensembles of frequencies by ${\cal E} := {\cal E}_S \cap {\cal E}_R$ and by $\omega$ the frequencies in it.  
   The term of first order is then simplified to
 \be
 - i \int_0^{\tau} {\rm d}u {\rm Tr}_R[\tilde{H}_{SR}(u), \rho_S^{Sc}(t)\rho_R^{Sc}(t)] = - i \lambda \tau \sum_{\omega \in {\cal E}}[\aso,\rho_S^{Sc}(t)] \bra A_R^{\dag}(\omega)\ket_{\rho_R^{Sc}(t)},
 \ee
 where $\bra {\cal O} \ket_{\rho} = {\rm Tr} {\cal O}\rho $ denotes the expectation value of the operator ${\cal O}$ taken in the state $\rho$.\\
 
 The second order term appearing on the right-hand side of \eqref{reduced} gives
 \bea
 - \int_0^{\tau}{\rm d}u_1 \int_0^{u_1}{\rm d}u_2 {\rm Tr}_R[\tilde{H}_{SR}(u_1),&[&\tilde{H}_{SR}(u_2),\rho_S^{Sc}(t)\rho_R^{Sc}(t)]]  \nn\\
  &&=-\sum_{\oms,\oms'} \gamma_{\oms,\oms'}(t,\tau) [\as A_S^{\dag}(\oms') \rho_S^{Sc}(t) - A_S^{\dag}(\oms')\rho_S^{Sc}(t)\as] + {\rm h.c.},
 \eea
  where
  \bea\label{gammaom}
  \gamma_{\oms,\oms'}(t,\tau) &=& \lambda^2 \sum_{\omr,\omr'}\bra \adr A_R(\omr')\ket_{\rho_R^{Sc}(t)} \int_0^{\tau} {\rm d}u_1 \int_0^{u_1}{\rm d}u_2 e^{i(\omr-\oms)u_1}e^{i(\omr'-\oms')u_2}\nn\\
 &=&  \lambda^2 \sum_{\omr,\omr'}\bra \adr A_R(\omr')\ket_{\rho_R^{Sc}(t)} \frac{i\tau}{\omr'-\oms'} e^{i(\omr-\oms-\omr'+\oms')\tau/2}\nn\\
  &&\hspace{2cm}\times\left[\sinc(\omr-\oms-\omr'+\oms')\tau/2 -e^{i(\omr'-\oms')\tau/2}\sinc(\omr-\oms)\tau/2\right]\nn\\
  &=& \frac{\tau^2\lambda^2}{2} \bra A_R^{\dag}(\oms) A_R(\oms')\ket_{\rho_R^{Sc}(t)}.
  \eea
   The last line was obtained using the convention established above. Then, for each $\oms$, there is only one $\omr$ such that $\omr \simeq \oms$ (meaning $|\omr - \oms|\tau \ll1$), and the others are such that $|\oms-\omr|\tau \gg 1$. It implies that for $\omr$ and $\omr'$ different from $\oms$ and $\oms'$,
   \be
  \Big| \int_0^{\tau} {\rm d}u_1 \int_0^{u_1}{\rm d}u_2 e^{i(\omr-\oms)u_1}e^{i(\omr'-\oms')u_2} \Big| \ll \frac{\tau}{|\omr'-\oms'|} \ll \tau^2,
  \ee
  yielding the last line of \eqref{gammaom}. As a side observation, we mention that the above approximation is equivalent to the secular approximation (it will appear more clearly after \eqref{repint}). Finally, the reduce dynamics of $S$ becomes
  \bea
  e^{i\tau H_S} \rho_{S}^{Sc}(t+\tau) &&e^{-i\tau H_S} = \rho_S^{Sc}(t) - i \lambda \tau \sum_{\omega \in {\cal E}}[\aso,\rho_S^{Sc}(t)] \bra A_R^{\dag}(\omega)\ket_{\rho_R^{Sc}(t)} \nn\\
 &&-\frac{\lambda^2\tau^2}{2}\sum_{\omega,\omega' \in {\cal E}} \bra A_R^{\dag}(\omega)A_R(\omega')\ket_{\rho_R^{Sc}(t)}[\aso A_S^{\dag}(\omega') \rho_S^{Sc}(t) - A_S^{\dag}(\omega')\rho_S^{Sc}(t)\aso] + {\rm h.c.}+ {\cal O}(\tau^3|\lambda|^3),\nn\\
 \eea
  where, as above, we denote by $\omega$ the (near-)resonant frequencies.\\
  
  To proceed it is convenient to go to the interaction picture, which we defined as $\rho_S(t) :=e^{iH_S t} \rho_S^{Sc}(t) e^{-iH_St}$. In this picture, the reduced evolution of $S$ for a time interval $\tau$ can be written as
 \be
 \rho_S(t+\tau) = (1+{\cal L}_{t,\tau} ) \rho_S(t),
 \ee
  where
  \bea
  {\cal L}_{t,\tau}\rho &=& - i \lambda \tau \sum_{\omega \in {\cal E}}[\aso,\rho] \bra A_R^{\dag}(\omega)\ket_{\rho_R^{Sc}(t)}e^{-i\omega t} \nn\\
 &&-\frac{\lambda^2\tau^2}{2}\sum_{\omega,\omega' \in {\cal E}} \bra A_R^{\dag}(\omega)A_R(\omega')\ket_{\rho_R^{Sc}(t)}e^{i(\omega'-\omega)t}[\aso A_S^{\dag}(\omega') \rho - A_S^{\dag}(\omega')\rho\aso] + {\rm h.c.}
 \eea
 
So far, we expressed the evolution of $S$ due to its interaction with $R$ for a duration $\tau$ (such that $\tau\gg |\omr-\oms|^{-1}$ for all $\omr \ne \oms$). We consider that the evolution of $S$ is made of a sequence of interactions with $R$ of duration $\tau_i$ separated by interval of free evolution of duration $T_i$. Denoting by $t_i$ the instant of time when $S$ and $R$ start interacting, (so that $t_{i+1} = t_i + \tau_i +T_i$), the state of $S$ at $t_n = t +\Delta t$ after $n$ such sequences is
\be
\rho_S(t + \Delta t) = (1+{\cal L}_{t_{n-1},\tau})...(1+{\cal L}_{t_{0},\tau})\rho_S(t)
\ee
where $t_0 = t$. Note that the intermediary free evolution does not affect the dynamics in interaction picture, $\rho_S(t_i + \tau_i +T_i) = \rho_S(t_i+\tau_i) = (1+{\cal L}_{t_i,\tau_i})\rho_S(t_i)$.
Since $\tau_i$ is much smaller than $\lambda^{-1}$ the action of the operators ${\cal L}_{t_{i},\tau_i}$ is small, $|{\cal L}_{t_i,\tau_i} \rho_S(t_i)| \ll 1$, so that 
\bea
\rho_S(t + \Delta t) &\simeq& (1+ {\cal L}_{t_{n-1},\tau_{n-1}} + ...+ {\cal L}_{t_{0},\tau_0})\rho_S(t)\nn\\
&=& \rho_S(t) - i \lambda  \sum_{\omega \in {\cal E}}[\aso,\rho_S(t)] \sum_{i=0}^{n-1}\tau_i\bra A_R^{\dag}(\omega)\ket_{\rho_R^{Sc}(t_i)}e^{-i\omega t_i} \nn\\
 &&\!\!\!\!\!\!\!\!\!\!\!\!\!\!-\lambda^2\sum_{\omega,\omega' \in {\cal E}} [\aso A_S^{\dag}(\omega') \rho_S(t) - A_S^{\dag}(\omega')\rho_S(t)\aso]\sum_{i=0}^{n-1}\frac{\tau^2_i}{2}\bra A_R^{\dag}(\omega)A_R(\omega')\ket_{\rho_R^{Sc}(t_i)}e^{i(\omega'-\omega)t_i} + {\rm h.c.}.\label{repint}
 \eea
The above expression can be simplified if we assume that $R$ is always prepared in the same state $\rho_R$. This brings us to the problem of the phase: coherences between levels of different energy accumulated a phase (even if $R$ is kept perfectly isolated). This phase depends on the ``history'' of each system, which appears in the terms $\bra A_R^{\dag}(\omega)\ket_{\rho_R^{Sc}(t_i)}$ and $\bra A_R^{\dag}(\omega)A_R(\omega')\ket_{\rho_R^{Sc}(t_i)}$. Then, unless $R$ is reinitialised with a well defined phase reference, or alternatively, identical systems $R$ are prepared and maintained coherent (which both are not happening in ``natural' systems and are highly demanding experimentally) the phases of the coherences are random.  
As a result, the unitary contribution (term of first order) in \eqref{repint} is null on average. Similarly, contributions from second order terms with $\omega \ne \omega'$ are also null on average. One should note that it turns out to be equivalent to assume that $R$ is prepared in states such that $\bra \ar \ket_{\rho_R} = 0$ and $\bra A_R^{\dag}(\omega)A_R(\omega')\ket_{\rho_R}= 0$ for $\omega \ne \omega'$ (which also corresponds to the stationarity condition). The above considerations provide a physical motivation for this assumption. We are thus left with
 \bea
\rho_S(t + \Delta t) &=& \rho_S(t) -n\frac{\lambda^2\tau^2}{2}\sum_{\omega \in {\cal E}}\bra A_R^{\dag}(\omega)A_R(\omega)\ket_{\rho_R} [\aso A_S^{\dag}(\omega) \rho_S(t) - A_S^{\dag}(\omega)\rho_S(t)\aso] + {\rm h.c.},
 \eea
 where $\tau^2:= \sum_{i=1}^{n-1}\tau_i^2/n$ denotes the average of the square of the interaction interval $\tau_i$.
 On average $S$ interacts $n=r\Delta t$ times with $R$ during a time interval $\Delta t$, where $r$ is the rate of repetition of the interactions. The coarse-grain time derivative of $\rho_S$ defined by $\dot{\rho}_S := \frac{1}{\Delta t}[\rho_S(t+\Delta t) - \rho_S(t)]$ (with $\Delta t$ such that $r   \tau^2\lambda^2\Delta t \ll 1$) is then 
 \be\label{L}
 \dot{\rho}_S = r {\cal L_{\tau}}\rho_S(t),
 \ee
 with the operator ${\cal L}_{\tau}$ can be rewritten in the from 
 \bea\label{LL}
 {\cal L_{\tau}}\rho_S(t) &=& \frac{\tau^2\lambda^2}{2} \sum_{\omega \in {\cal E}} \bra A_R(\omega)A_R^{\dag}(\omega)\ket_{\rho_R}  \left\{[A_S(\omega)\rho_S, A^{\dag}_S(\omega)] + [A_S(\omega),\rho_S A_S^{\dag}(\omega)]\right \}.
 \eea
This the master equation indicated in the main text with 
\be\label{gammacm}
\Gamma(\omega) := \frac{r\tau^2 \lambda^2}{2}{\rm Tr}_R[\rho_R A_R(\omega)A_R^{\dag}(\omega)].
\ee
 Importantly, we recall that the ladder operators of $S$ and $R$ appearing in \eqref{LL} are the sum of the ladder operators of the quasi degenerate Bohr frequencies. One should note that from Eq. \eqref{thermaleq} of Section \ref{apptempth} the relation used in the main text, 
\be\label{thide}
G(-\omega) = e^{-\omega/T_R}G(\omega),
\ee
 comes straightforwardly ($\hbar=1$, $k_{B}=1$).
 
 We summarise in the following the four conditions used to derived the reduced dynamics \eqref{L} and \eqref{LL}. Firstly, the condition on which relies the essence of the collisional model, namely $R$ is reinitialised after each interaction with $S$ or equivalently, identical systems prepared in the same state are sent successively to interact with $S$. 
 Secondly, the condition $\tau_i \ll |\lambda|^{-1}$ for all $i$, where $\tau_i$ is the duration of the $i^{\rm th}$ interaction and plays the same role as the bath correlation time in dissipation driven by baths. Therefore, the condition $\tau_i \ll |\lambda|^{-1}$ is equivalent to $\tau_i \ll T_R$, where $T_R$ is the evolution timescale of $S$ (in the interaction picture), which justifies the Born Markov approximation in bath-induced dissipation \cite{Cohen_Book}. The third condition is that the phase of the prepared state $\rho_R^{Sc}(t_i)$ is random. Finally, the last condition is the existence of resonant or nearly resonant frequencies of $S$ and $R$ such that $|\omega_S-\omega_R| \tau_i \ll 1$ while the other frequencies are far from resonance, namely $|\omega_S-\omega_R| \tau_i \gg 1$. One should note that combined with the first condition it means that detuning as large as $|\lambda|$ between $\omega_S$ and $\omega_R$ still satisfies the near resonant condition. 
 As a concluding observation, this last condition plays the role of the secular approximation. Alternatively, we can derive the same dynamics \eqref{LL} without this last condition. In doing so, after applying the third condition, one ends up with  
 \bea
\rho_S(t + \Delta t) &\simeq& (1+ {\cal L}_{t_{n-1},\tau} + ...+ {\cal L}_{t_{0},\tau})\rho_S(t)\nn\\
&=& \rho_S(t) -\lambda^2\sum_{\omega_S,\omega_S' \in {\cal E}_S} [A_S(\omega_S) A_S^{\dag}(\omega_S') \rho_S(t) - A_S^{\dag}(\omega_S')\rho_S(t)A_S(\omega_S)]\sum_{\omega_R\in{\cal E}_R}  \bra A_R^{\dag}(\omega_R)A_R(\omega_R)\ket_{\rho_R}\nn\\
 &&\hspace{4.5cm}\times \int_0^{\tau}du_1\int_0^{u_1}du_2 e^{i(\omega_R-\omega_S)u_1}e^{i(\omega_R-\omega_S')u_2}\sum_{i=0}^{n-1}e^{i(\omega_S'-\omega_S)t_i} + {\rm h.c.}.
 \eea
 One should note that in the above expression we assumed for simplicity the equal duration of each interaction with $R$, $\tau_i = \tau$. The more general situation can be treated in a similar way.
If the time intervals $T_i$ of free evolution are of random length, then the instants of time $t_i$ are random variables and thus $\sum_{i=0}^{n-1}e^{i(\omega_S'-\omega_S)t_i}$ is a sum of random phases which cancels out on average (unless $\omega_S=\omega_S'$). However, if the time intervals $T_i$ are regular, $T_i=T$ for all $i$, then $\sum_{i=0}^{n-1}e^{i(\omega_S'-\omega_S)t_i} = e^{i(\omega_S'-\omega_S)t}\frac{1-e^{i(\omega_S'-\omega_S)\Delta t}}{1-e^{i(\omega_S'-\omega_S)(\tau+T)}}$. For $n=\Delta t/(\tau+T)\gg1$, the term $ e^{i(\omega_S'-\omega_S)t}\frac{1-e^{i(\omega_S'-\omega_S)\Delta t}}{1-e^{i(\omega_S'-\omega_S)(\tau+T)}}$ is much smaller than $n$ if $\omega_S\ne\omega_S'$ so that one can retain only terms with $\omega_S=\omega_S'$. Alternatively, one can invoke the secular approximation, namely the phase $e^{i(\omega_S'-\omega_S)t}$ is oscillating rapidly while $\rho_S(t)$ is evolving on a timescale $(r\tau^2\lambda^2)^{-1}$ so that the contributions from terms with $\omega_S\ne\omega_S'$ can be neglected assuming $|\omega_S'-\omega_S|^{-1} \ll (r\tau^2\lambda^2)^{-1}$. One finally obtains
\bea
\dot{\rho}_S = r {\cal L}_{\tau}\rho_S(t) = r\sum_{\omega_S \in {\cal E}_S} \Gamma_{\tau}(\omega_S)  \left[ A_S(\omega)\rho_S(t)A_S^{\dag}(\omega) - A_S^{\dag}(\omega)A_S(\omega)\rho_S(t)\right] + {\rm h.c.},
\eea
with $\Gamma_{\tau}(\omega) = \lambda^2\sum_{\omega_R\in{\cal E}_R}  \bra A_R(\omega_R)A_R^{\dag}(\omega_R)\ket_{\rho_R} \int_0^{\tau}du_1\int_0^{u_1}du_2 e^{-i(\omega_R-\omega_S)(u_1+u_2)}$, which is a generalisation of \eqref{gammacm}.

\section{Thermal baths}\label{thbath}
We consider here that $S$ is interacting with a bath $R$ in an unitary way through the following Hamiltonian,
\be\label{htot}
H_{Tot} = H_S + H_R + H_{SR},
\ee
where $H_S$ and $H_R$ are the free Hamiltonians of $S$ and $R$, respectively. Importantly, $S$ can be composed of a single system or many non-interacting subsystems.
The coupling term is of the form $H_{SR} = \lambda P_S P_R$, $P_S$ and $P_R$ are operators of $S$ and $R$, respectively, and $\lambda$ is a real coupling constant. We consider that $S$ has a discrete spectrum, and denoting by $\epsilon$ its eigenvalues and by $\pi_{\epsilon}$ the projector onto the eigenspace associated with $\epsilon$ we can decompose $P_S$ in a sum of eigenoperators of $H_S$ \cite{Petruccione_Book},
\be\label{decomppr}
P_S = \sum_{\omega} A_S(\omega),
\ee
where the ladder operators 
\be\label{ladderop}
A_S(\omega):= \sum_{\epsilon' -\epsilon = \omega} \pi_{\epsilon} P_S \pi_{\epsilon'},
\ee
 satisfy the commutation relation $[H_S,A_S(\omega)]=-\omega A_S(\omega)$, (setting $\hbar=1$) as mentioned in the main text. The sum runs over the Bohr frequencies $\omega$ of $S$.
Furthermore, we assume that $R$ is in a stationnary state, $[\rho_R,H_R]=0$, and that $S$ and $B$ are interacting weakly so that the Born and Markovian approximations are legitimate. Following the usual procedure a Gorini-Kossakowski-Lindblad-Sudarshan (GKLS) quantum master equation \cite{gklsme} for the reduced dynamics of $S$ is derived \cite{Petruccione_Book},
\bea\label{mer}
\dot{\rho}_S &=&  \sum_{\omega} \Gamma(\omega) \left[A_S(\omega)\rho_S A^{\dag}_S(\omega) -A_S^{\dag}(\omega) A_S(\omega)\rho_S\right ] + {\rm h.c.},
\eea
where $\rho_S$ is the density operator of $S$ in the interaction picture (with respect to $H_S$) and $\dot{\rho}_S$ its time derivative. The complex function $\Gamma(\omega)$ is given by 
\be\label{gamma}
\Gamma (\omega) := \lambda^2 \int_0^{\infty} ds e^{i\omega s}{\rm Tr}[\rho_R P^I_R(s) P_R],
\ee
with $P^I_R(s) = e^{iH_R s}P_R e^{-iH_R s}$ is the bath coupling operator in the interaction picture.  
Importantly, \eqref{mer} is the same master equation as Eq. (3) of the main text with the expression of $\Gamma(\omega)$ given above \eqref{gamma}. 
We also defined the bath spectral density as $G(\omega):= \Gamma(\omega) + \Gamma^*(\omega) =\lambda^2 \int_{-\infty}^{\infty} ds e^{i\omega s}{\rm Tr}[\rho_R P^I_R(s) P_R]$ which also satisfies \eqref{thide} (proof in Section IV). 
In the main text we mention that the ratio $G(\omega)/G(-\omega)$ can be related to the apparent temperatures of $R$ defined by $e^{\omega/{\cal T}_R(\omega)} = \frac{\bra A_R(\omega)A_R^{\dag}(\omega)\ket_{\rho_R}}{\bra A_R^{\dag}(\omega)A_R(\omega)\ket_{\rho_R}}$. To see this we consider that we can decompose the coupling operator $P_R$ in eigenoperators in the form $P_R = \sum_{\omega} A_R(\omega) $. For instance for a bosonic bath composed of a collection (rigorously a continuum) of modes and with coupling operator of the form $P_R=\sum_k (g_k^*b_k + g_k b_k^{\dag})$, this would amount to define $A_R(\omega) := \sum_k g_k^* b_k \delta(\omega-\omega_k)$ when $\omega > 0$ (and $A_R(\omega):= g_k b_k^{\dag} \delta(\omega+\omega_k)$ when $\omega <0$), where $\omega_k$ is the energy associated to the mode $k$.
Computing the expression of the bath spectral density one obtains
\bea
G(\omega) &=&2\pi \lambda^2 \sum_k \big[ g_k \bra b_k^{\dag} P_R\ket_{\rho_R}\delta(\omega+\omega_k) + g_k^* \bra b_k P_B\ket_{\rho_R}\delta(\omega -\omega_k) \big]\nn\\
&&=2\pi \lambda^2\sum_{k,k'} \big[ g_kg_{k'}^*\bra b_k^{\dag}b_{k'}\ket_{\rho_R}\delta(\omega+\omega_k) \delta(\omega+\omega_{k'}) +  g_k^*g_{k'}\bra b_kb_{k'}^{\dag}\ket_{\rho_R}\delta(\omega-\omega_k)\delta(\omega -\omega_{k'}) \big]
\eea
where the second line was obtained using the sationarity hypothesis $[\rho_R,H_R]=0$ wich implies that $ \bra b_k^{\dag}b_{k'}\ket_{\rho_R}=0$ unless $\omega_k = \omega_{k'}$. For $\omega >0$ we have $G(\omega) =  2\pi \lambda^2\sum_{k,k'} g_k^*g_{k'}  \bra b_kb_{k'}^{\dag}\ket_{\rho_R}\delta(\omega-\omega_k)\delta(\omega -\omega_{k'}) =2\pi \lambda^2 \bra A_R(\omega)A_R^{\dag}(\omega)\ket_{\rho_R}$.
It follows the announced identity 
\be
\frac{G(\omega)}{G(-\omega)} = \frac{\bra A_R(\omega)A_R^{\dag}(\omega)\ket_{\rho_R}}{\bra A_R^{\dag}(\omega)A_R(\omega)\ket_{\rho_R}}.
\ee

\section{Apparent temperature for arbitrary transition coefficients}
For arbitrary transition coefficient $\alpha_{n-1,n,g,g'} := \bra n-1,g|\rho_X|n,g'\ket$ the apparent temperature takes the following form
\bea
&&\frac{\omega}{{\cal T}_X} = \nn\\
&&\log\frac{\sum_{n=1}^N\left[\sum_{g=1}^{l_{n-1}}\bra n-1,g|\rho_X |n-1,g\ket\sum_{g'=1}^{l_n}|\alpha_{n-1,n,g,g'}|^2+ \sum_{g,j}^{l_{n-1}}\bra n-1,j|\rho_X|n-1,g\ket\sum_{g'=1}^{l_n}\alpha_{n-1,n,g,g'}\alpha_{n-1,n,j,g'}^{*}\right]}{\sum_{n=1}^N\left[\sum_{g=1}^{l_{n}}\bra n,g|\rho_X |n,g\ket\sum_{g'=1}^{l_{n-1}}|\alpha_{n-1,n,g',g}|^2+ \sum_{g,j}^{l_{n}}\bra n,g|\rho_X|n,j\ket\sum_{g'=1}^{l_{n-1}}\alpha_{n-1,n,g',g}\alpha_{n-1,n,g',j}^{*}\right]},\nn\\
\eea
which can be simplified to the expression Eq. (10) of the main text when all coefficient $\alpha$ are equal. Although more involved, the above expression maintains the same meaning and consequence as in the main text. Namely, the first term of both numerator and denominator are the contributions from the population weighted by the transition coefficients. The second term of both numerator and denominator are the contributions from the coherence between degenerate levels, weighted again by the transition coefficients. As a consequence, the value of the apparent temperature can be affected dramatically by the coherences. Interestingly, the non-uniformity of the values of the transition coefficients can even amplifies the effect of the coherences. A more detailed discussion about the consequences of such phenomenon can be found in the main text. \\

\section{Apparent temperatures for thermal states}\label{apptempth}
When $X=R$, $S$ is in a thermal state at temperature $T_X$, $\rho_X = Z^{-1}e^{-H_X/T_X}$, where $Z$ is the partition function, one can derive the following identity,
\bea\label{thermalid}
e^{H_X/T_X}A_X(\omega)e^{-H_X/T_X} &=& \sum_{n=0}^{\infty} \frac{1}{n!}\frac{1}{T^n_X}{\rm Ad}^n_{H_X} A_X(\omega) \nn\\
&=& \sum_{n=0}^{\infty} \frac{1}{n!}\frac{1}{T^n_X}(-\omega)^n A_X(\omega)\nn\\
&=& e^{-\omega/T_X}A_X(\omega),
 \eea
 where ${\rm Ad}^n_{M}N := [M, {\rm Ad}^{n-1}_M N]$ and ${\rm Ad}^0_M N = N$. We recall that the ladder operators are such that $[H_X,A_X(\omega)] = -\omega A_X(\omega)$ so that ${\rm Ad}^n_{H_X} A_X(\omega) = (-\omega)^n A_X(\omega)$.
From \eqref{thermalid} one obtains,
\bea\label{thermaleq}
\langle A_X^{\dag}(\omega) A_X(\omega) \rangle_{\rho_X} &=& {\rm Tr}\big[Z^{-1}e^{-H_X/T_X}A_X^{\dag}(\omega) A_X(\omega) \big]\nn\\
&=&  Z^{-1}{\rm Tr}\big[A_X^{\dag}(\omega)e^{-H_X/T_X}e^{H_X/T_X} A_X(\omega)\nn\\
&&\hspace{4cm}\times e^{-H_X/T_X} \big]\nn\\
&=&  Z^{-1}{\rm Tr}\big[A_X^{\dag}(\omega)e^{-H_X/T_X}e^{-\omega/T_X} A_X(\omega) \big]\nn\\
&=& e^{-\omega/T_X} \langle A_X(\omega) A_X^{\dag}(\omega) \rangle_{\rho_X} .
\eea
Substituting in the expression of the apparent temperature (Eq. (6) of the main text), one obtains ${\cal T}_X(\omega) = T_X$ for all Bohr frequencies $\omega$. Importantly, this also shows that the function $G(\omega)$ (see previous Sections I and \ref{thbath}) satisfies  the thermal identity \eqref{thide} as soon as $\rho_R$ is a thermal state.\\

\section{Apparent temperature of the phaseonium}
In Eq. (5) of \cite{Scully_2003}, the average photon number of the cavity $n_{\phi}$ is given to follow the dynamics
\be
\dot{n}_{\phi} = \alpha [2\rho_{aa}(n_{\phi}+1) - (\rho_{bb}+\rho_{cc} + \rho_{bc}+\rho_{cb})n_{\phi}],
\ee
 where  $\rho_{xy}:=\langle x|\rho_R|y\rangle$, $x,y=a,b,c$, and $\alpha$ is a simple rate factor. The steady state average photon number is obtained by taking $\dot{n}_{\phi} =0$ in the previous equation, leading to
 \be\label{nphi}
\frac{n_{\phi}+1}{n_{\phi}} = \frac{\rho_{bb}+\rho_{cc} + \rho_{bc}+\rho_{cb}}{2\rho_{aa}}.
\ee
 The steady state temperature $T_{\phi}$ of the cavity is related to the average photon number by $n_{\phi} = \left(e^{\Omega/k_B T_{\phi}}-1\right)^{-1}$, where $\Omega$ is the frequency of the cavity. Substituting in \eqref{nphi} one obtains
 \be
 k_BT_{\phi} = \Omega \left(\ln{\frac{\rho_{bb}+\rho_{cc}+\rho_{bc}+\rho_{cb}}{2\rho_{aa}}}\right)^{-1},
 \ee
 which is precisely the expression obtained in Eq. (13) of the main text (substituting $\Omega$ by $\omega$). One should note that in order to serve better the purpose of the article the expression of $T_{\phi}$ used in \cite{Scully_2003} is an approximated version of the above expression.

\section{Hotter or colder apparent temperature thanks to coherence}
The apparent temperature is hotter thanks to the coherences if and only if $1/{\cal T}_X \leq 1/{\cal T}_X^0$ (${\cal T}_X^0$ is the apparent temperature without coherence), which implies from Eq. (10) of the main text,
\be
\frac{\sum_{n=1}^{N} l_n(\rho_{n-1} + c_{n-1})}{\sum_{n=1}^{N} l_{n-1}(\rho_n + c_n)} \leq \frac{\sum_{n=1}^{N} l_n\rho_{n-1}}{\sum_{n=1}^{N} l_{n-1}\rho_n }.
\ee
This can be rewritten in the form
\be
{\cal C}^+ \geq {\cal C}^- e^{-\omega/{\cal T}_X^0},
\ee
where ${\cal T}_X^0 := \omega \left[\ln {\frac{\sum_{n=1}^{N} l_n\rho_{n-1}}{\sum_{n=1}^{N} l_{n-1}\rho_n }}\right]^{-1}$, ${\cal C}^+ := \sum_{n=1}^{N} l_{n-1} c_n$, and ${\cal C}^-:=\sum_{n=1}^{N} l_n c_{n-1}$.
For the three-level system in the $\Lambda$-configuration considered in the main text, the apparent temperature is hotter thanks to the coherence if and only if $c_0$ is negative, which can also be verified directly from Eq. (13) of the main text. 
It is worth noting that we use the expression ``hotter" in a broad sense (a negative temperature is considered hotter than a positive temperature).

\section{Hotter and colder apparent temperature thanks to correlations}
From Eq. (11) of the main text, the apparent temperature is left hotter thanks to correlations if and only if 
\be
\frac{ \sum_{i=1}^m \langle A_i A_i^{\dag}\rangle_{\rho_R}+c }{\sum_{i=1}^m \langle A_i^{\dag} A_i \rangle_{\rho_R} + c} \leq \frac{ \sum_{i=1}^m \langle A_i A_i^{\dag}\rangle_{\rho_R} }{\sum_{i=1}^m \langle A_i^{\dag} A_i \rangle_{\rho_R} },
\ee
which is equivalent to the condition 
\be
c  (e^{\omega/{\cal T}_X^0} -1)\geq 0,
\ee
where ${\cal T}_X^0 := \omega \left[\ln{ \frac{ \sum_{i=1}^m \langle A_i A_i^{\dag}\rangle_{\rho_R} }{\sum_{i=1}^m \langle A_i^{\dag} A_i \rangle_{\rho_R} }}\right]^{-1}$ is the apparent temperature without correlations. 
If ${\cal T}_X^0 \geq 0$ (which happens when the individual subsystems are not in inverted-population states), the apparent temperature is hotter (colder) thanks to the correlations if and only if $c$ is positive (negative). If ${\cal T}_X^0 \leq 0$ (for instance the individual subsystems are in inverted-population states) then it is the contrary: the apparent temperature is hotter (colder) thanks to the correlations if and only if $c$ is negative (positive).
As in the previous Section, we use the expression ``hotter" in a broad sense (a negative temperature is considered hotter than a positive temperature).

\section{Steady state of an indistinguishable pair of two-level atoms}
We consider the dynamics described by Eq. (14) of the main text for an indistinguishable pair of atoms,
\bea\label{indsm}
\dot{\rho}_S^I &=& -i\Omega_L \sum_{i=1}^2 [\sigma_i^{+}\sigma_i^{-},\rho_S^I]-i\Omega_{1,2}[\sigma_1^{+}\sigma_{2}^{-}+\sigma_1^{-}\sigma_{2}^{+},\rho_S^I] \nn\\
&&+ g [n(\omega) +1] (2S^-\rho_S^I S^+ -S^+S^-\rho_S^I - \rho_S^IS^+S^-)\nn\\
&&+ g n(\omega) (2S^+\rho_S^I S^- -S^-S^+\rho_S^I - \rho_S^IS^-S^+),\nn\\
\eea
where $n(\omega)$ is the bath mean excitation number at the frequency $\omega$, $\sigma_i^+$ and $\sigma_i^-$ are the ladder operators of the atom $i$ at the position $\vec r_i$, $S^+ = \sum_{i=1}^2\sigma_i^+$ and $S^-= \sum_{i=1}^2\sigma_i^- $ are the {\it collective} ladder operators of the pair of atoms, $\Omega_L$ is the Lamb shift, $g$ is the effective coupling between the atoms and the field mode (see expression in the main text), and the term proportional to $\Omega_{1,2}$ corresponds to the Van der Waals interaction between the two atoms (see expression in the main text).
 The dynamics can be easily solved by considering the basis $\{|\psi_0\ket,|\psi_+\ket,|\psi_-\ket,|\psi_1\ket\}$ with $|\psi_{\pm}\ket =(|01\ket\pm|10\ket)\sqrt{2}$, $|\psi_{0}\ket = |00\ket$, and $|\psi_1\ket=|11\ket$. In such basis the collective ladder operators can be expressed as $S^{+}=\sqrt{2}|\psi_+\ket \bra\psi_0|+\sqrt{2}|\psi_1\ket\bra\psi_+|$ and $S^{-}=\sqrt{2}|\psi_0\ket \bra\psi_{+}|+\sqrt{2}|\psi_{+}\ket\bra\psi_1|$. From \eqref{indsm} one obtains the following dynamics for the populations $p_i:=\bra \psi_i|\rho_S|\psi_i\ket$, $i=0,1,+,-$, 
 \bea
 &&\dot{p}_1 = 4 g n(\omega)p_+ - 4g[n(\omega)+1]p_1\nn\\
 &&\dot{p}_0=4g[n(\omega)+1]p_+ -4gn(\omega)p_0\nn\\
 &&\dot{p}_+ = 4g[n(\omega)+1](p_1-p_+)-4gn(\omega)(p_0-p_+)\nn\\
 &&\dot{p}_- = 0.
 \eea 
Using the fact that $\dot{p}_1+\dot{p}_0+\dot{p}_+ = 0$, one can easily find the steady state populations to be $p_0^{\infty}=Z_{\rm ind}^{-1}$, $p_+^{\infty}=Z_{\rm ind}^{-1}e^{-\omega \beta_B}$ and $p_1^{\infty}= Z_{\rm ind}^{-1}e^{-2\omega \beta_B}$ with $Z_{\rm ind} = 1 +e^{-\omega\beta_B}+e^{-2\omega\beta_B}$, and $\beta_B$ is the inverse temperature of the bath. For the coherences, defined as $\rho_{ij}:=\bra \psi_i|\rho_S^I|\psi_j\ket$, $i,j \in \{0,1,+,-\}$, one obtains (including the Lamb shift to the interaction picture),
\bea 
&&\dot{\rho}_{+,-} = -2\big[g(2n(\omega)+1) + i\Omega_{1,2} \big] \rho_{+,-}\nn\\
&&\dot{\rho}_{1,-} = -\big[2g(n(\omega)+1)+i\Omega_{1,2}\big]\rho_{1,-} \nn\\
&&\dot{\rho}_{0,-} = -\big[2gn(\omega)+i\Omega_{1,2}\big]\rho_{0,-} \nn\\
&&\dot{\rho}_{1,0} = -2g(2n(\omega)+1)\rho_{1,0}
\eea
which straightforwardly gives $0$ as steady state solution. The dynamics of the two remaining coherences is coupled,
\bea
&&\dot{\rho}_{1,+} = -\big[2g(3n(\omega)+2)-i\Omega_{1,2}\big]\rho_{1,+} + 4gn(\omega)\rho_{+,0}\nn\\
&&\dot{\rho}_{+,0} =  -\big[2g(3n(\omega)+1)+i\Omega_{1,2}\big]\rho_{+,0} + 4g(n(\omega)+1)\rho_{1,+},
\eea
and also leads to $0$ as steady state solution. Finally, one can write the steady state solution in the form,
\be
  \rho_{\rm ind}^{\infty} = Z_{\rm ind}^{-1} \Big(e^{-2\omega\beta_B}|\psi_1\ket\bra \psi_1|+e^{-\omega\beta_B}|\psi_+\ket\bra\psi_+| + |\psi_0\ket\bra\psi_0|\Big),
\ee
as announced in the main text.

\end{widetext}

\end{document}